# From Wurtzite Nanoplatelets to Zinc Blende Nanorods: Simultaneous Control of Shape and Phase in Ultrathin ZnS Nanocrystals

Liwei Dai,[a] Rostyslav Lesyuk,[a,b] Anastasia Karpulevich,[a] Abderrezak Torche,[a] Gabriel Bester,[a] Christian Klinke[*a,c,d]

[a] *Institute of Physical Chemistry, University of Hamburg, Martin-Luther-King-Platz 6, 20146 Hamburg, Germany*
[b] *Pidstryhach Institute for applied problems of mechanics and mathematics of NAS of Ukraine, Naukowa str. 3b, 79060 Lviv & Department of Photonics, Lviv Polytechnic National University, Bandery str. 12, 79000 Lviv, Ukraine*
[c] *Department of Chemistry, Swansea University – Singleton Park, Swansea SA2 8PP, UK*
[d] *Institute of Physics, University of Rostock, Albert-Einstein-Straße 23, 18059 Rostock, Germany*

**Abstract:** Ultrathin semiconductor nanocrystals (NCs) with at least one dimension below their exciton Bohr radius receive a rapidly increasing attention due to their unique physicochemical properties such as strong quantum confinement, large surface-to-volume ratio, and giant oscillator strength. These superior properties highly depend on the shape and crystal phase of semiconductor NCs. Slight changes in the shape and phase of NCs can cause significant changes in their properties. Therefore, it is crucial to controllably synthesize semiconductor NCs. Here, we demonstrate not only the synthesis of robust well-defined ultrathin ZnS nanoplatelets (NPLs) with excitonic absorption and emission, but also the precise shape and phase control of ZnS NCs based on a soft template strategy. The key feature of our approach is the tuning of the sulfur precursor amount, resulting in a simultaneous shape/phase transformation between wurtzite (WZ) ZnS NPLs and zinc blende (ZB) ZnS nanorods (NRs) at moderate temperatures (150 °C). UV-vis absorption and photoluminescence (PL) spectra reveal very distinct optical properties between WZ-ZnS NPLs and ZB-ZnS NRs. UV-vis absorption spectra of WZ-ZnS NPLs clearly exhibit a sharp excitonic peak that is not observed in ZB-ZnS NRs. Besides, the PL characterization shows that WZ-ZnS NPLs have a narrow excitonic emission peak (292 nm), while the ZB-ZnS NRs exhibit a broad collective emission band consisting of four emission peaks (335, 359, 395, and 468 nm). The appearance of excitonic features in the absorption spectra of ZnS NPLs is explained by interband electronic transitions, which is simulated in the framework of density functional theory (DFT). The presented simple and effective synthetic strategy opens a new path to synthesize further NCs with shape and phase control for advanced applications in electronics and photonics.

* Corresponding author: christian.klinke@swansea.ac.uk

**Introduction**

After the discovery of graphene,[1-2] ultrathin semiconductor nanocrystals (NCs) with at least one dimension below their exciton Bohr radius, typically in the sub-5 nm regime,[3] have attracted intensive research interests due to strong quantum confinement effects, large surface-to-volume ratio, giant oscillator strength, and potential applications in optoelectronic devices, sensors, energy conversion and storage, and catalysis.[4-6] These advantages are largely based on the existence of the ultrathin size in at least one dimension, resulting in strong quantum confinement.[7-8] Among many ultrathin semiconductor NCs, quasi-two-dimensional (2D) nanoplatelets (NPLs) with atomically precise thicknesses have been the most studied ones. Such semiconductor NPLs exhibit unique behavior such as narrow spectral linewidths in absorption and emission as well as large absorption cross-sections because of their well-controlled size in one dimension.[7] So far, numerous studies on CdSe NPLs have been reported by several groups.[9-12] Compared to Cd-based NCs, zinc sulfide (ZnS) exhibits very low toxicity. In addition, ZnS, as an important wide-band gap semiconductor, has a large direct band gap of approximately 3.7-3.8 eV at room temperature and a small Bohr radius of 2.5 nm, which makes it a promising candidate for electroluminescence, sensor, laser technology, and photocatalysis.[13-14] By colloidal method, several zinc chalcogenide quasi-2D structures have been already synthesized in recent years. In 2011, Karan et al. presented both wurtzite (WZ) and zinc blende (ZB) ZnS NPLs synthesis assisted by $Mn^{2+}$ introduction involving oleylamine (OAm).[15] Later Park et al. showed ZnSe NPLs with narrow excitonic absorption peak at room temperature using OAm/octylamine (OTA) mixture as the solvent and ligand.[16] Similar soft-template method has been adapted by Buffard et al. for synthesizing WZ ZnS NPLs by low temperatures.[17] Eventually, Bouet et al. showed principal possibility to form ZB ZnS NPLs by sequential cation exchange from CdS template.[18] However no excitonic photoluminescence (PL) from reported and mentioned structures was showed yet. To the best of our knowledge, there are only few reports on colloidal synthesis of ultrathin ZnS NPLs, and the synthesis of robust well-defined structures with excitonic emission is still a challenge, especially in terms of simultaneous shape/phase transition.

Shape control is highly significant in the fabrication of semiconductor NCs because of their shape-dependent optical and electrical characteristics.[19-22] In particular, semiconductor NCs can be tailored to gain optimized properties via a shape control strategy. A variety of approaches including thermal evaporation,[23] vapor liquid-solid processes,[24] and thermolysis[25] have been used to controllably synthesize ZnS NCs. In recent years, the wet chemical method was employed to modify and manipulate the shape of ZnS NCs because of distinct advantages like low-cost, easy operation, and solution-processibility. Wang et al. and Liu et al. reported a one-pot solvothermal preparation of ZnS nanostructures from zero-dimensional (0D) nanoparticles to one-dimensional (1D) NWs in the presence of ethylenediamine (EN) as a soft template.[26-27] Acharya et al. reported on the synthesis of shape controlled ZnS NCs designed into nanodots (NDs), NRs, and NWs by varying the precursor concentration, capping ligands, and annealing temperature.[28] Nonetheless, among these works, the growth mechanisms responsible for the formation of various ZnS nanostructures are still not clear due to complex interdependencies in synthesis and growth conditions. More detailed experiments are needed to corroborate the relevant inferences.

Phase control is another key factor in the manipulation of optoelectronic properties of semiconductor NCs. As is well known, ZnS generally exits as two types of structural polymorphs, the cubic ZB and the hexagonal WZ structure having a direct wide band gap of 3.77 and 3.72 eV, respectively.[29-31] Due to differences in atomic arrangement, the properties of ZB-ZnS are quite different from that of WZ-ZnS. For instance, the ZB-ZnS and WZ-ZnS exhibit very different surface chemical activities and electronic characteristics because of the different Zn- and S- terminated surfaces.[32] Recently, the WZ phase of ZnS was shown to produce a spontaneous polarization and internal electric field stimulating better charge carriers separation and transfer and making the WZ phase attractive for more efficient photocatalytic processes and energy conversion.[33] In previous reports, ZnS NCs synthesized by a wet chemical method were usually of ZB (sphalerite) structure, which is a low-temperature phase.[34] Wurtzite is the high-temperature polymorph of ZnS, which can be formed at temperatures around 1023 °C.[35] Although the temperature for the synthesis of WZ-ZnS NCs has been reduced to 150°C,[36] it is still difficult to achieve the phase transition at low temperature. In already mentioned work, Karan et al. achieve reversible ZB/WZ phase changes in ZnS nanostructures at 300 °C by reversible insertion/ejection of doping ions induced by a thermocyclic process.[15] However, the obtained ZnS NCs were not pure and the used temperature is still relatively high. Thus, to achieve the phase transition in pure ZnS NCs at low temperature by using wet chemical methods still remains a big challenge.

In the present work, we report a simultaneous control of shape and phase in pure ultrathin ZnS NCs at relatively low temperature (150 °C) based on a soft template strategy. The colloidal rectangular WZ-ZnS NPLs with ultra-small thickness (1.81 ± 0.20 nm) were synthesized with a proper amount of sulfur precursor. By changing the amount of sulfur precursor, the shape/phase transformation between WZ-ZnS NPLs and ZB-ZnS NRs can be achieved, which was systematically investigated by transmission electron microscope (TEM), X-ray diffraction (XRD), selected area electron diffraction (SAED), high-resolution (HR) TEM, and Fast Fourier Transformation (FFT). UV-vis absorption spectra of the samples reveal quite distinguished optical properties between WZ-ZnS NPLs and ZB-ZnS NRs arising from the quantum confinement. Additionally, we show the excitonic PL of synthesized WZ NPLs at room temperature. The mechanism for the formation of ultrathin ZnS NCs was supported by TEM, small-angle XRD, and $^1$H NMR. To give explanations to the observed absorption and emission spectra, we perform *ab initio* calculations of the band-structure within the framework of DFT and simulation of the dipole transitions for the ZnS NPLs.

**Results and Discussions**

In a first series of experiments, the amount of sulfur was varied while keeping all other parameters constant. TEM images (Figure 1A-F) clearly show the shape evolution of ZnS NCs upon increase of the amount of sulfur. With a small amount (0.45 mmol) of sulfur, rectangular ZnS NPLs with lateral dimension of 7.1 ± 0.80 nm and 19.4 ± 2.19 nm (Figure S1A) were synthesized. The thickness of the NPLs is 1.81 ± 0.20 nm (Figure S2A), directly measured by TEM on NPLs that stand vertically on the TEM grid. Based on the structural parameters of ZnS NCs (Figure S2B), the NPLs correspond to a ZnS slab with ca. 5 monolayers (MLs) along the thickness direction (1 ML corresponds to one crystal unit of WZ ZnS). By increasing the amount of sulfur, the shape of ZnS NCs changes from NPLs (0.45 mmol, Figure 1A) to a mixture of NPLs and NRs (0.90, 1.35, and 2.70 mmol, Figure 1B-D) to NRs (4.05 mmol, Figure 1E). The average diameter of the NRs is 2.84 ± 0.32 nm (Figure S1B) measured from TEM images. At a larger amount (8.10 mmol) of sulfur, NRs with a diameter of 3.62 ± 0.43 nm (Figure S1B) were obtained (Figure 1F). According to the above experimental results, a small amount of sulfur is advantageous for the formation of ZnS NPLs, but the minimum required amount of sulfur is around 0.45 mmol. If the amount of sulfur is further decreased to 0.30 or 0.15 mmol, only small ZnS nanoparticles with irregular shape were synthesized (Figure S3A and B).

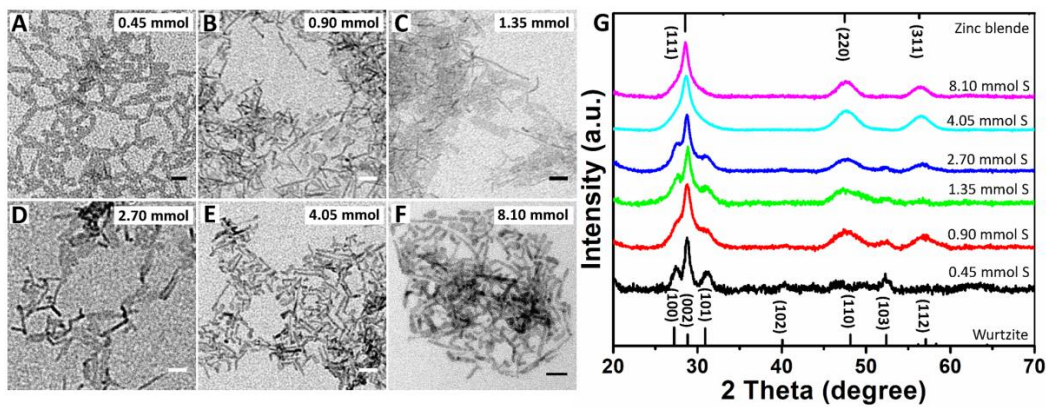

**Figure 1.** TEM images of ultrathin ZnS NCs synthesized with different amounts of sulfur. (A) ZnS NPLs (0.45 mmol), (B-D) a mixture of ZnS NPLs and NRs (0.90, 1.35, and 2.70 mmol), (E-F) ZnS NRs (4.05 and 8.10 mmol). The scale bars represent 20 nm.(G) XRD patterns of ZnS NCs synthesized with different amounts of sulfur. At the bottom (wurtzite, ICPDS 00-080-0007) and the top (zinc blende ICPDS 00-077-2100), the diffractograms of the bulk ZnS are shown.

X-ray diffraction (XRD) was performed to determine the crystal structure of ZnS NCs. The XRD patterns (Figure 1G) indicate that a change in the crystal structure from WZ to ZB occurred as the shape of ZnS NCs varied from NPLs to NRs. At a small amount (0.45 mmol) of sulfur, the characteristic diffraction peaks correspond to the WZ structure (ICPDS 00-080-0007). With increasing the amount of sulfur, the hexagonal (100), (101), (102), and (103) peaks disappear and the characteristic peaks of cubic (111), (220), and (311) planes corresponding to the ZB structure of ZnS became discernable, revealing the crystal structure changes from WZ to ZB. When a large amount (4.05 and 8.10 mmol) of sulfur was used, the XRD pattern displays all characteristic peaks of the bulk ZB-ZnS pattern (ICPDS 00-077-2100) without impurity peaks. Liu and co-workers showed that the width of the peaks is correlated with the growth direction for the ZB-ZnS NWs.[37] A narrow peak is assigned to long axis directions, while a broad peak is related to short axis directions. Thus, the narrowest (111) peak and the broadest (220) peak of NRs reveal the long axis growth in [111] direction and the short axis growth in [220] direction. Moreover, it is noticeable that the (220) peak corresponding to the diameter of NRs gradually narrowed with the increase in sulfur precursor, indicating that the average crystallite sizes were growing larger. The mean diameters of NRs synthesized with 4.05 and 8.10 mmol S were 2.99 and 3.63 nm respectively, as estimated from Scherrer's equation by fitting the (220) XRD peak, which is consistent with the sizes of NRs determined by the TEM observations.

The selected area electron diffraction (SAED) pattern (Figure 2A, E) supports the findings of the crystal structure change from WZ to ZB when the shape of ZnS NCs varied from NPLs to NRs. Figure 2B shows a high-resolution (HR) TEM image of an individual ZnS NPL where the well-resolved lattice fringe pattern illustrates that the NPL is well crystallized. The observed lattice spacings were calculated to be 3.09 Å and 3.27 Å, matching fairly well the (0001) and (10-10) plane spacings of the bulk WZ-ZnS structure (ICPDS 00-080-0007). The corresponding Fast Fourier Transformation (FFT) (Figure 2C) clearly reveals the lateral orientation of ZnS NPLs in [0001] and [10-10] directions. The energy-dispersive X-ray spectroscopy (EDS) spectrum (Figure S4) shows the atomic ratio of Zn:S to be 1:0.85, confirming the composition of the as-synthesized product is ZnS. The atomistic models for WZ-ZnS NPLs (Figure 2D) show that side (0001) facets are either Zn-rich (grey) or S-rich (yellow), while the lateral (1-210) facets and side (10-10) facets exhibit a mixed composition of Zn and S. The HRTEM image (Figure 2F) reveals a well-resolved lattice structure of an individual ZnS NR, indicating that the NR is single-crystalline. The distinct lattice spacings were measured to be approximately 3.13 Å, corresponding to the (111) plane spacing of the bulk cubic ZB-ZnS structure (ICPDS 00-077-2100). Besides, the HRTEM image (Figure 2F) shows that the [111] crystallographic axis is parallel to the long axis of the NR. This result indicates that the NRs grew along the [111] direction (Figure 2H), a thermodynamically favorable growth direction,[37] which is consistent with the XRD observations (Figure 1G). Additional structural confirmations are provided by the FFT pattern (Figure 2G) taken from the corresponding HRTEM image. A comparison of the FFT patterns of NPLs (Figure 2C) and NRs (Figure 2G) reveals that the ZnS NPLs show multiple spots corresponding to different orientations of the crystallographic planes, while the ZnS NRs show only (111) spots for the preferential growth direction. The EDS spectrum (Figure S5) suggests that the NRs are composed of Zn and S and the atomic ratio of Zn:S is 0.95:1, which is in agreement with the stoichiometric ratio of ZnS compound. The atomistic models for ZB-ZnS NRs (Figure 2H) show that the (111) facets are either Zn-rich (grey) or S-rich (yellow), while the side {110} and {11-2} facets are mixtures of Zn and S.

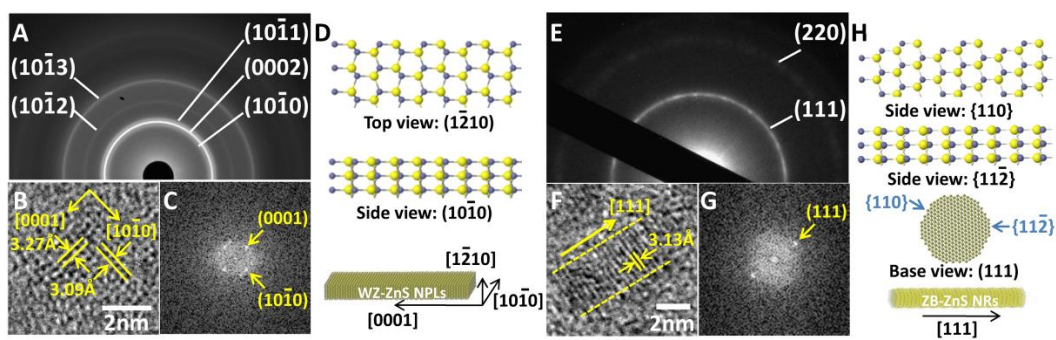

**Figure 2.** Characterization of the ZnS NPLs and NRs. (A, E) SAED pattern, (B, F) HRTEM images, and (C, G) the corresponding FFT pattern of the ZnS NPLs and NRs synthesized with 0.45 and 4.05 mmol sulfur, respectively. (D, H) Schematic illustration of the crystallographic nature of the ZnS NPLs and NRs. Grey: Zn atoms; Yellow: S atoms.

The optical properties of ZnS NPLs and NRs were characterized by steady-state UV-vis absorption and PL spectroscopy. The absorption spectrum of the WZ-ZnS NPLs (Figure 3A and B) shows a distinct narrow peak at λ=283 nm (4.38eV) with FWHM of about 164 meV followed by less pronounced peaks at 4.8 eV and 5.62 eV with a shoulder at 5.78 eV. The emission intensity is centered at 292 nm (4.25 eV) with FWHM of 105 meV revealing a Stokes-shift of about 9 nm (130 meV).

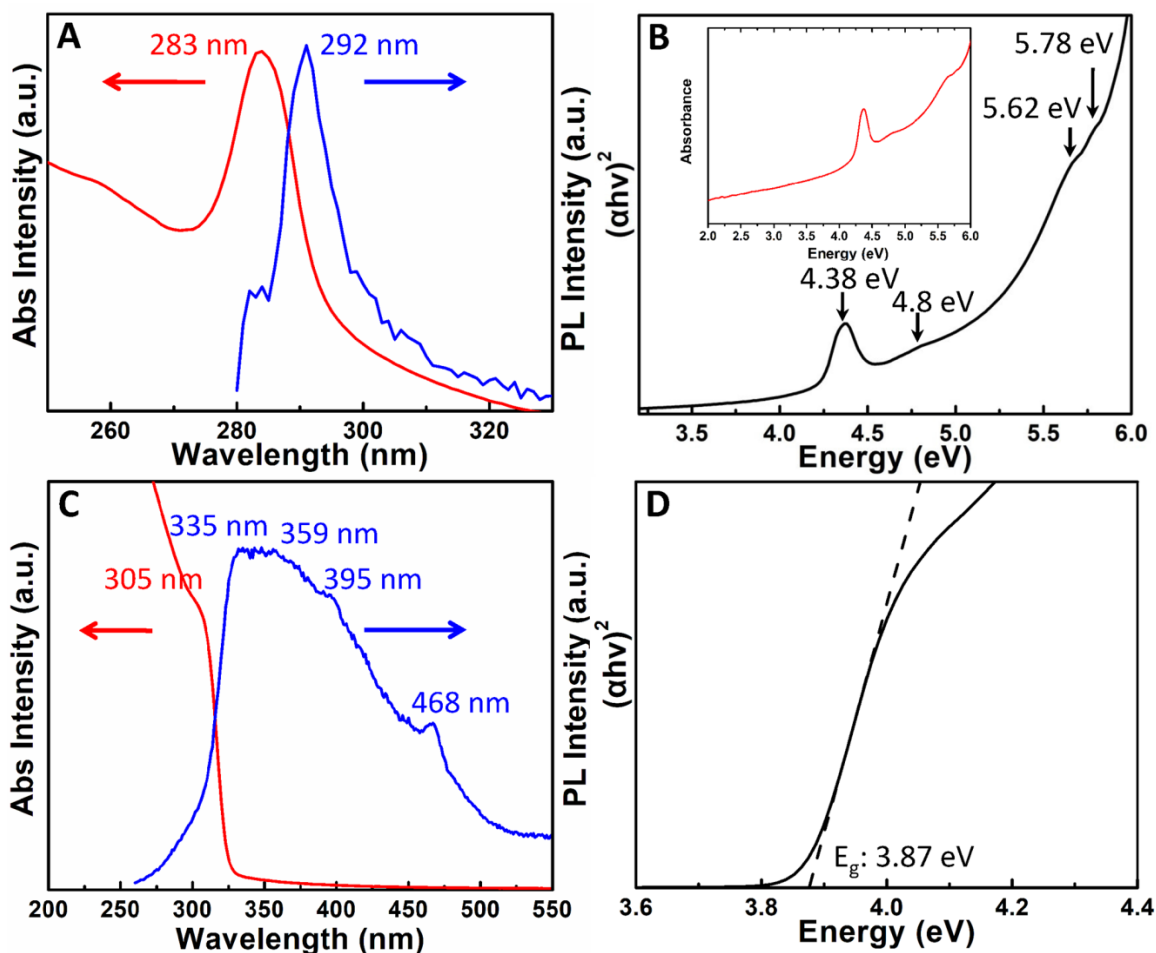

**Figure 3.** UV-vis absorption spectra (red line) and photoluminescence (blue line) of ZnS NCs. (A) ZnS NPLs synthesized with 0.45 mmol sulfur and (C) ZnS NRs synthesized with 4.05 mmol sulfur. (B) and (D): $(\alpha h\nu)^2$ vs photon energy representation of the absorption spectra for ZnS NPLs and NRs. The inset in (B) shows the normalized absorbance of NPLs for a wider range of energies revealing the scattering background. The band gap ($E_g$) of ZnS NRs are analyzed from the UV-vis absorption spectra by calculating the absorption coefficient ($\alpha$). $E_g$ is calculated by using the relation for direct bandgap semiconductors: $(\alpha h\nu)^2 = E_g - h\nu$.

We attribute the first absorption peak of the NPLs to an excitonic transition substantially enhanced by the Coulombic attraction between electrons and holes. The sharpness of the absorption peak suggests that the NPLs are homogeneous in thickness (one population), which agrees with the TEM results (Figure S2). The appearance of several absorption features in the absorption spectra of NPLs, and generally in 2D systems, was attributed to electronic transitions from heavy (hh)-, and light-hole (lh) levels to the first conduction band,[7] as well as interband transitions to higher conduction sub-bands.[38] For colloidal CdSe NPLs the main absorption feature was shown to originate from hh-e and lh-etransitions,[7] separated by ca. 180 meV for the ZB[9] and 220 meV for the WZ NPLs[39]. On the other hand, the splitting of the absorption bands based on the level structure of the top of the valence band was shown to be smaller for bulk ZnS films[40] because of lighter atoms and hence weaker spin-orbit coupling. Pronounced hydrogenic1s, 2s lines were observable even at room temperature, the (so)-e transition smeared out after heating over 200 K. However, the spectral separation of these bands was in the range of 70 meV. To undoubtedly resolve the question of the origin of the absorption features in our spectra, we performed *ab initio* calculations (see Methods section)[41-45] of the WZ ZnS slab with [11-20] growth direction and 1.91 nm width (5 MLs), which corresponds to the average width of the NPLs measured with TEM. From the theoretical simulation of the dipole transitions using Γ-point wave functions, we can conclude that the first absorption peak corresponds to a sum of three transitions, namely from the nodeless states derived from the A-, B- and C-bulk valence (WZ) bands to the nodeless conduction band minimum state (CBM). In Figure 4 we show the experimental absorption spectrum and the calculated dipole allowed transitions (panel A) along with the eigenvalues and eigenfunctions (panel B). The observed first absorption peak is dominated by the third valence band state (VBM-2), which has an offset from the VBM less than 100 meV. The dominance of the VBM-2 state in the absorption peak obviously contributes to the observed considerable Stokes shift. Our calculations show that crystal field, spin-orbit and confinement effects on the valence band splitting for the 2D ZnS NPLs with experimental width are in the range of tens of meV, and hence not large enough to produce experimentally distinct absorption peaks. The second absorption peak at around 4.8 eV corresponds to a transition between conduction and valence band states with a single node ($h_4$-$e_1$ transition, between second sub-band states). The third absorption peak at around 5.62 eV corresponds to a transition between nodeless states ($h_1$-$e_3$ transition), while the electron state $e_3$ originates from the second bulk conduction band at the Gamma-point. The weak shoulder corresponds to transitions between states with two nodes ($h_5$-$e_2$, from the second sub-band). The agreement between the calculations and the experimental results is qualitatively good with some quantitative differences. The second absorption peak is slightly too high in energy and the lowest peak slightly too low. Note that our single particle gap is at 4.25 eV, which is over 100 meV below the experimental result. We attribute the differences to the lack of correlation effects in the calculations and possibly the presence of out-of-plane strain which would explain our underestimated band gap. Indeed, due to surface reconstruction, the interatomic distances may be altered leading to the appearance of strain and to a shift of bands.[46] Analyzing the XRD data, we note that the (002) reflex corresponding to the lateral direction exactly coincides with the reference. At the same time the (10-10) and (10-11) which contain information from directions piercing the NPLs' flat basal plane at an angle different from zero are shifted to higher angles, which might be attributed to the presence of compressive strain in the thickness direction. In any case, the large blue shift (ca. 0.6 eV) relative to the bulk WZ-ZnS band gap (3.77 eV)[31, 47] indicates the existence of a very strong one dimensional (1D) quantum confinement effect,[48] which is mainly attributed to the ultrathin thickness of the NPLs. The present results explain the relatively large broadening of the first absorption band in comparison with the PL and absorption broadening of CdSe NPLs, since it consists in our case of unresolved hh-e, lh-e and so-e contributions ($h_{0,1,2}$-$e_0$ transitions).

The absorption spectrum of the ZB-ZnS NRs (Figure 3C) shows a more traditional shape consisting of a continuum-like absorption with a shoulder at 305 nm, which can be attributed to the band edge absorption onset. Figure 3D shows that the band gap value of ZnS NRs estimated by the Tauc linearization is approximately 3.87 eV, which is slightly blue-shifted by 0.15 eV compared to the bulk ZB-ZnS band gap (3.72 eV),[30, 47] implying a weak quantum confinement effect in agreement with recent *ab-initio* simulations of ZnS NRs.[49] The PL spectrum of ZnS NRs exhibits an inhomogeneously broadened non-symmetric band consisting of peaks at 335 nm and 359 nm and weak shoulders at 395 nm and 468 nm, respectively. These peaks can be attributed to the band-edge emission (335 nm, 3.7 eV), the excitonic emissions at interstitial sulfur and the interstitial zinc (359 nm and 395 nm), and dangling sulfur bonds at the interface of ZnS (468 nm).[50-51] In the current case, the well-known deep-trap emission (around 450 nm) generated from surface sulfur vacant sites is not observed, which can be explained by the excess sulfur in the system as demonstrated by the EDS data (Figure S5).[50, 52]

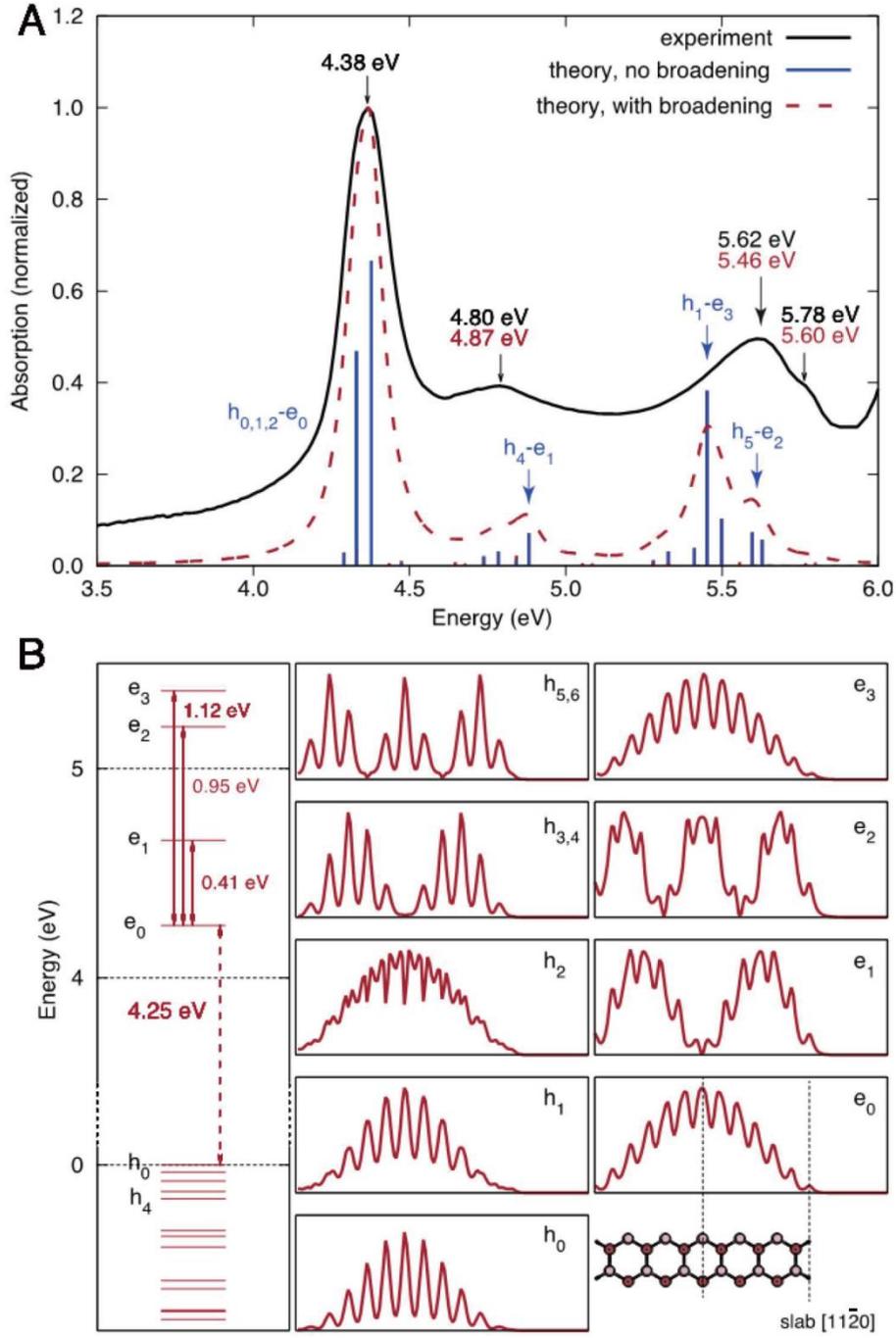

**Figure 4.** (A) Normalized absorption spectrum (black) of ZnS WZ NPLs after subtraction of a scattering background along with calculated dipole transitions (red dashed). The first exciton energy was fitted to the experimental value and a temperature broadening applied to the results (see Methods). (B) Single particle eigenvalues at Γ-point (left) and 1d atomistic wave functions (right) of the relevant near-band-gap states (see Supplementary). Atomic structure of the ZnS slab used in the simulations (bottom right).

To investigate the formation mechanism of the ultrathin ZnS NPLs, we studied the $ZnCl_2$-OTA-OAm complex in regard to a soft template mechanism.[53-55] It has been reported that the inorganic metal halide and the alkylamine can form a lamellar structure by the van der Waals attraction between hydrocarbon sidechains of the alkylamine.[56-58] In a recent work, Buhro et al. reported the synthesis of ultrathin (1 nm) PbS NPLs by using analkylamine-based soft template strategy.[59] The formation of a lamellar, amine-bilayer mesophase was confirmed, which well explains the two-dimensional growth of PbS NCs. In this present reaction system, a mixture of OTA and OAm was chosen as the soft template, which plays two important roles in the formation of well-separated ZnS NPLs: 1) the optimum reactivity of the OTA; 2) the long OAm molecular chain can serve to enhance the steric repulsion between organic layers, resulting in weakened interactions between the NPLs.[3, 56] The complex was precipitated by adding acetone to the mixture of OTA, OAm, and $ZnCl_2$, followed by centrifugation. $^1$H NMR (Figure S6) confirms the presence of OTA and OAm ligands in the $[ZnCl_2(OTA, OAm)_x]$ (x: the total number of N atoms in OTA and OAm molecules coordinated with Zn atoms) mesostructure. A TEM image (Figure 5A) of white solid (Inset: $[ZnCl_2(OTA, OAm)_x]$ complex in hexane) collected upon precipitation shows the formation of the bundled strands. The weak contrast of the strands in the TEM image indicates that they are mainly composed of organic species. A small-angle powder X-ray diffraction (SAXRD) pattern (Figure 5B) was recorded from the precipitate, confirming a typically lamellar structure of the compound. The series of XRD peaks can be assigned to '00$l$' ($l$=1, 2, 3, 4...) with a layer spacing (d) of ~4.56 nm, which is close to but smaller than the double length (4.87 nm) of the OAm molecule chain (the molecule length of 2.436 nm). The length of OAm molecule chains was calculated according to the formula, L (nm) = 0.15 + 0.127n,[60] where n is the number of carbon atoms in the alkyl chain. The slight difference (ca. 0.31 nm) between the layer spacing and the double length of the OAm molecule chain might be attributed to two main reasons. On the one hand, note that this value (2.436 nm) is for the case where the alkyl chain of the OAm molecule is fully expanded,[60-61] while the real value of the OAm length should be smaller than 2.436 nm due to the shrinkage of the molecule chain. On the other hand, the interpenetration (η) of OAm molecule chains results in the reduction of the interlay distance. Such an interpenetration structure of the OAm bilayer has been reported in previous works for CdSe nanobelts.[62]

As we have shown, the shape evolves from NPLs to NRs with increasing amount of sulfur precursor (Figure 1A-F). Peng and co-workers showed that the shape-evolution was correlated with the monomer concentration.[63-64] For a given solution system, a high monomer concentration can support the growth of the nanocrystal in its 1D growth stage, which generates rod-shaped NCs. At a low monomer concentration, the dot-shaped NCs are preferential formed due to the 3D-growth stage. In our case, we clearly found that the ZnS NRs were synthesized with a high sulfur concentration compared to NPLs. Both the HRTEM image (Figure 2F) and the FFT pattern (Figure 2G) confirm the 1D growth stage. Thus, for the formation of NRs, it can be attributed to the high monomer concentration.[65] The overall synthetic procedure of ultrathin ZnS NPLs is presented in Figure 5C. In the first step, a [$ZnCl_2$ (OTA, OAm)$_x$] lamellar mesostructure is formed by the reaction of $ZnCl_2$, OTA, and OAm at 100 °C. The newly formed [$ZnCl_2$ (OTA, OAm)$_x$] lamellar mesostructure then can be used as a soft template for the synthesis of ultrathin ZnS NPLs under the template-guided effects.

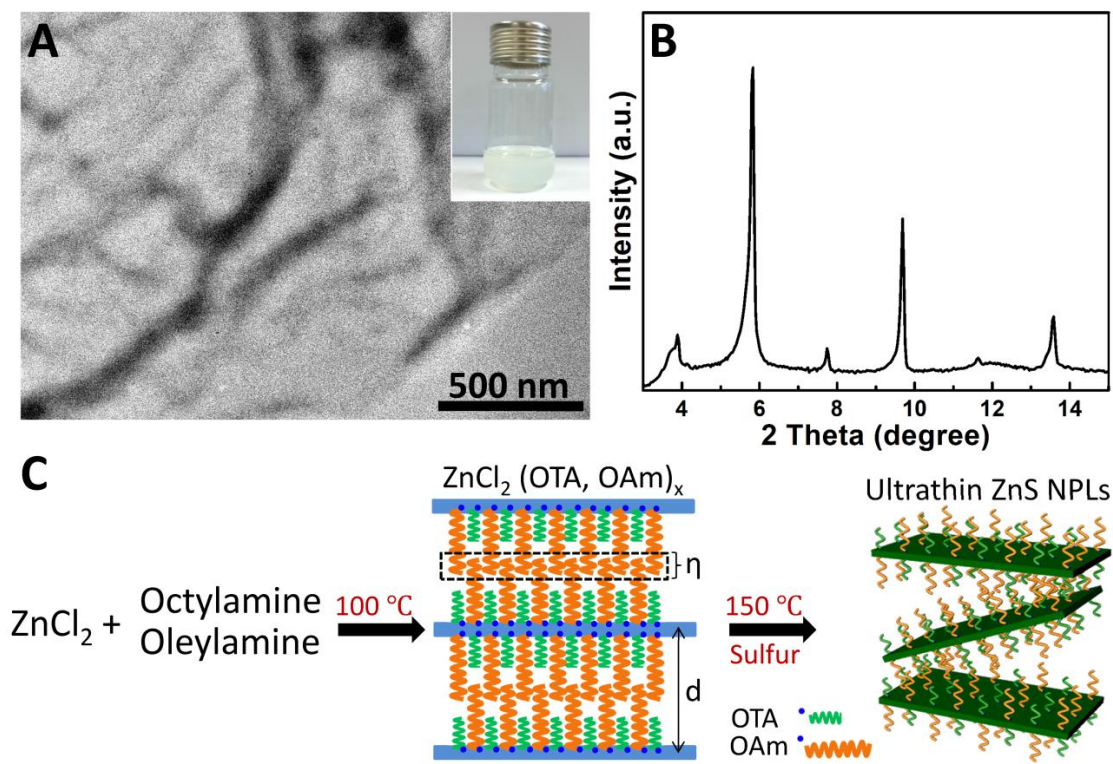

**Figure 5.** TEM and SAXRD characterization of the [$ZnCl_2$ (OTA, OAm)$_x$] complex. (A) TEM image and photograph (inset) of the [$ZnCl_2$ (OTA, OAm)$_x$] strands formed from octylamine, oleylamine, and $ZnCl_2$. (B) SAXRD pattern for the polymer strands. (C) Schematic illustration of the synthesis of ultrathin ZnS NPLs.

Additionally, the different volume ratios (R = $V_{OAm}$/$V_{OTA}$) between OAm and OTA were found to play important roles in obtaining various morphological ZnS NCs. The TEM images (Figure S7) clearly show the shape transformation of ZnS NCs upon varying R. An R = 2/1 is essential for the preparation of uniform ultrathin ZnS NPLs (Figure S7C). With R more than 2/1, small irregular ZnS nanoparticles were synthesized (Figure S7A and B). When R = 1/1 was used as the mixed solvent, ZnS NPLs formed together with a large amount of nanoparticles (Figure S7D and Figure S8). If R is less than 1/1, aggregated ZnS nanosheets with different contrast in the TEM images (Figure S7E-G) were observed and the aggregation is probably due to a lack of steric repulsion of the OAm chains. The XRD patterns (Figure S9) of the ZnS NCs obtained at different R show all characteristic peaks of the bulk WZ-ZnS pattern (ICPDS 00-080-0007). The volume-dependent study demonstrates that both OTA and OAm take part in the assembly of the soft template needed for the formation of homogeneous NPLs, indirectly confirming that the double-lamellar structure is responsible for the 2D formation of ZnS. However, the phase choice (WZ) was not dependent on R. This evidence clearly supports the exceptional role of the sulfur precursor amount in the nucleation and growth process in different crystal phases of ZnS.

In conclusion, we demonstrated simultaneous shape and crystal phase control in ultrathin ZnS NCs at relatively low temperature (150 °C) based on a soft template strategy. The colloidal rectangular WZ-ZnS NPLs with excitonic absorption and emission were synthesized with a proper amount of sulfur precursor. By changing the amount of sulfur precursor, the shape was tuned from NPLs to NRs and the phase was transformed from WZ to ZB. The shape/phase changes can be attributed to changes in nucleation and growth of NCs induced by different reactant concentration. The synthesized ZnS NCs exhibit distinct optical properties, especially the excitonic transition and emission feature in WZ-ZnS NPLs, which is not reported so far. The observed features in the absorption spectrum at room temperature demonstrate the genuine quasi-2D nature of our ZnS NPLs. An adequate explanation to the observed optical features such as a considerable Stokes shift and several excitonic peaks in the absorption spectrum is given by the help of *ab-initio* calculations. We find that observed absorption and emission bands arise from electronic interband transitions ($h_{0,1,2}$–$e_0$, $h_4$–$e_1$, $h_1$–$e_3$, $h_5$–$e_2$), and the splitting of hh-, lh- and so-sublevels contributes merely to the broadening of excitonic transitions observed at room temperature. The shape/phase controlled evolution shown in this work endows the NCs with tunable optoelectronic properties, making them highly attractive for optoelectronic and catalytic applications.

**Author Information**
Corresponding Author
*E-mail: christian.klinke@swansea.ac.uk
ORCID
Christian Klinke: 0000-0001-8558-7389
Liwei Dai: 0000-0002-4360-6075
Notes

The authors declare no competing financial interest.

**Acknowledgements**

The authors gratefully acknowledge financial support of the European Research Council via the ERC Starting Grant "2D-SYNETRA" (Seventh Framework Programme FP7, Project: 304980) and the China Scholarship Council (CSC), PRC. C.K. thanks the German Research Foundation DFG for financial support in the frame of the Cluster of Excellence "Center of ultrafast imaging CUI" and the Heisenberg scholarship KL 1453/9-2. We thank Eugen Klein, Stefan Werner, Almut Barck, and Daniela Weinert (Universität Hamburg) for help with the sample measurements.

**Table of Content**

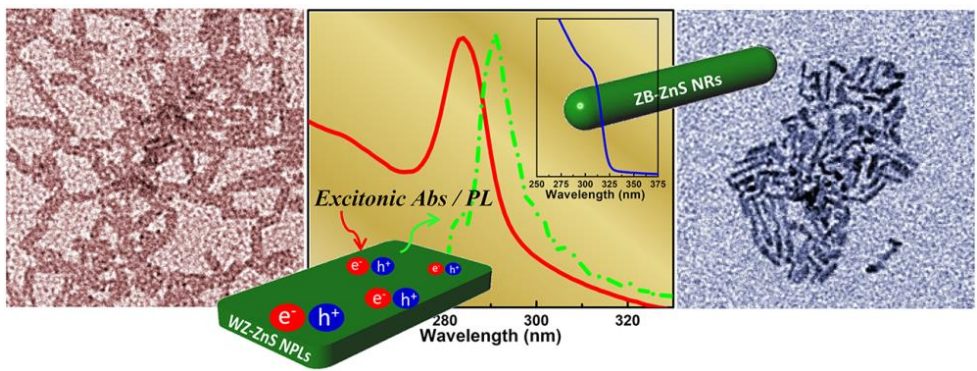